
\hsize= 16.0 cm
\vsize= 22.0 cm
\nopagenumbers                          
\overfullrule 0 pt
\hoffset=0.20 cm
\voffset=0.5 cm
\def\doublespace{\baselineskip=\normalbaselineskip 
\multiply\baselineskip by 2}

\doublespace

\def\EXPRA{1}
\def\EXPRB{2}
\def\EXPRC{3}
\def\EXPRD{4}
\def\SCOTT{5}
\def\STEIN{6}
\def\BARKH{7}
\def\PRIES{8}
\def\IEEEREV{9}
\def\AHARO{10}
\def\HOWAR{11}
\def\GROUP{12}
\def\TEBBL{13}
\def\CRAIK{14}

\def\dbar{d\!\!\! {}^{-}}

\noindent Magnetic Particles

\noindent M.~A.\ Novotny \hfil\break
Supercomputer Computations Research Institute \hfil\break
Florida State University\hfil\break
Tallahassee, FL 32306-4052, U.S.A. \hfil\break
and \hfil\break
Department of Electrical Engineering \hfil\break
2525 Pottsdamer Street \hfil\break 
Florida A\&M University -- Florida State University\hfil\break
Tallahassee, FL 32310-6046, U.S.A. \hfil\break

\noindent P.~A.\ Rikvold \hfil\break
Center for Materials Research and Technology \hfil\break
and \hfil\break
Department of Physics\hfil\break
Florida State University\hfil\break
Tallahassee, FL 32306-3016, U.S.A. \hfil\break
and \hfil\break
Supercomputer Computations Research Institute \hfil\break
Florida State University, Tallahassee, FL 32306-4052, U.S.A. \hfil\break

\vfill
\eject

\centerline{Introduction}

Ever since the development of the magnetic compass revolutionized 
navigation in the second century A.D., applications of magnetic materials 
have become essential in most branches of engineering.  In many of these 
applications the magnetic materials are utilized in the form of microscopic 
magnetic particles.  Some modern representative examples from electrical 
and electronics engineering are 
nonvolatile storage of information on magnetic tapes and disks, 
magnetic inks, 
refrigerators that make use of the magnetocaloric effect, 
ferrofluid vacuum seals, 
and the microscopic machines known as micro-electromechanical systems (MEMS).  
It is likely that these current and emergent engineering applications of 
microscopic and nanoscopic magnetic particles will be joined by many more 
in the next few decades.  

Even though 
magnetic particles have been used for a very long time, 
our understanding of their physical behaviors 
is relatively new and still quite limited.  
In addition, the current understanding of the physical properties 
of magnetic particles, 
and the engineering applications that are made possible by these properties, 
cannot easily be described without sophisticated 
mathematics.  
Some of the reasons for this are listed below.  \hfil\break
1)~~The magnetism of magnetic particles is 
fundamentally quantum-mechanical in origin.  Hence, 
understanding the properties of magnetic particles 
requires the use of quantum mechanics, either 
directly 
or by the inclusion of 
quantum-mechanical effects into phenomenological models.  \hfil\break
2)~~Magnetism involves vector quantities, which have 
a direction as well as a magnitude.  
The magnetic field ${\vec H}$ (given in units of A/m), 
the magnetic moment per unit volume of a magnetic substance 
${\vec M}$ (given in units of A/m), 
and 
the magnetic induction or magnetic flux density 
${\vec B}$ (given in units of Wb/m$^2$ or Tesla) are all vector 
quantities, and their 
interrelationships must be expressed in terms of vector 
and tensor equations.  \hfil\break
3)~~For ferromagnetic particles, which are the ones used in 
most engineering applications, the relationships between 
the three vector fields ${\vec H}$, ${\vec B}$, and 
${\vec M}$ depend on the history of the particle --- 
how the field and magnetization have varied in the past, and possibly 
how the particle was manufactured.  \hfil \break
4)~~Engineering applications of magnetic particles are determined by 
physical properties that depend on energetic 
effects which originate from a number of different physical mechanisms. 
For example, in spherical particles 
the coercive field (defined below) is due primarily to 
the energy associated with crystalline anisotropy, while in elongated 
needle-like single-domain particles 
the dipole-dipole interactions (the magnetostatic energy) 
are the most important in determining the coercive field.  
\hfil\break
Commonly these complications 
are what make magnetic particles so useful in engineering applications.  
For example, the dependence of ${\vec M}$ on the detailed history of 
the particle is what makes magnetic recording possible.  

Most engineering applications utilize large numbers of 
magnetic particles, which may differ in size, orientation, 
composition, etc.  
However, a detailed 
understanding of the behavior of such 
an assortment of particles requires knowledge of 
the properties of individual particles.  
The properties of systems consisting of 
large numbers of particles can be obtained by an appropriate averaging 
over the size, orientation, shape, type, and location 
of individual particles.   

This article will focus on single magnetic particles. 
It is only in very recent years that it has become 
possible to study the behavior of individual particles experimentally. 
This is due to the development of 
better methods to produce well-characterized magnetic particles and 
to the development of measurement techniques 
with resolutions at the nanometer scale.  
These ultrahigh resolution techniques include
magnetic force microscopy (MFM)
({\EXPRA}), 
micro-SQUID devices ({\EXPRB}), 
Lorentz transmission electron microscopy ({\EXPRC}), 
and
giant magnetoresistive (GMR) measurements ({\EXPRD}).  

\centerline{Types of Magnetic Materials}

At the atomic level, a magnetic material is an arrangement of 
local magnetic moments.  
Fundamentally, each such magnetic moment is 
a quantum-mechanical quantity 
arising from either the intrinsic spin or 
the orbital motions of the electrons 
of an atom or molecule.  
Each of these localized magnetic moments, 
often called spins for brevity, 
may be thought of as a small bar magnet.  
The microscopic structures of the most widely 
studied types of magnetic materials are shown schematically in Fig.~1.  
This figure shows 
the lowest-energy arrangement (the ground state) 
of the local magnetic moments or spins
in (a) ferromagnetic, (b) antiferromagnetic, and (c) ferrimagnetic 
materials.  
The total magnetizations due to the spin arrangements 
shown in Fig.~1 
can be found by adding the vectors together.  
In a ferromagnet the spins add up to a large vector. 
For an antiferromagnet they add up to zero.  
For a ferrimagnet there is some cancellation, 
but the spins still add up to a nonzero value.  

The alignment of the spins in Fig.~1 (a)--(c) correspond to  
a temperature of absolute zero.  At nonzero temperatures, 
the alignment of the spins is somewhat random due to thermal 
fluctuations.  
At temperatures above a critical temperature, 
$T_{\rm c}$ (which is different for different materials), 
the thermal fluctuations are so large that the 
spin arrangement becomes essentially random, 
the total magnetization is zero, and the material becomes paramagnetic 
(as shown in Fig.~1d).  
For ferromagnets in zero applied magnetic field 
the magnetization vanishes at $T_{\rm c}$ and 
remains zero above $T_{\rm c}$.  
Since most engineering applications of magnetic materials 
use ferromagnets, the remainder of this article will 
focus on ferromagnetic particles.  

\centerline{Relationships Between the Vector Fields}

The fundamental relationship between the vector fields is given by 
$$
{\vec B} = \mu_0 \left ( {\vec H} + {\vec M} \right )
\eqno(1)
$$
where $\mu_0= 4\pi\times10^{-7}$~H/m is the permeability of free 
space.  Equation~(1) is true for all materials, even for nonlinear 
ones.  

For a linear homogeneous and isotropic material the 
relationship between ${\vec M}$ and ${\vec H}$ is linear and given by 
$$
{\vec M} = \chi_{\rm m} {\vec H} 
\eqno(2)
$$
where $\chi_{\rm m}$ 
is a material dependent quantity called the magnetic 
susceptibility.  In this case it is possible to write 
$$
{\vec B} 
= \mu_0\left[1+\chi_{\rm m}\right] {\vec H} 
= \mu_0\mu_{\rm r}{\vec H} = 
\mu {\vec H}
\eqno(3)
$$
where $\mu$ is the permeability of the medium and the parameter 
$\mu_{\rm r}$ is its relative permeability.

In ferromagnetic materials, however, the relationship between the 
three vector fields is generally nonlinear and history dependent.  
Thus simple relationships 
such as those of Eq.~(2) and Eq.~(3) are not justified.  In this case 
it is necessary to talk about a hysteresis loop.  
Figure~2 shows a hysteresis loop, with identification of 
the coercive field $H_{\rm c}$, the remanent (spontaneous) magnetization 
$M_{\rm r}$, and the saturation magnetization $M_{\rm s}$ 
(the maximum magnetization of a magnetic particle in a strong field).  
The area of the hysteresis loop corresponds to the work which must be 
done in taking the magnetic material through one cycle of the 
applied field.  This work is converted to heat 
and represents a major source of energy loss in devices such as 
transformers and motors.  
It was first studied systematically by 
C.~P.\ Steinmetz (\STEIN). 

One useful classification of magnetic materials for engineering purposes 
is into soft and hard magnets.
Soft magnetic materials are usually used for
cores of transformers, generators, and motors, and the heads in 
magnetic tape and disk devices; applications 
which require a low coercive field, $H_{\rm c}$, small hysteresis 
loop area to minimize heat generation, or high permeability.  
Hard magnetic materials are used for  
electric sensors, loudspeakers, electric meters, magnetic 
recording media, and other 
uses that require a high coercive field, a 
high remanence, or a large hysteresis loss.  

Not all hysteresis loops have the shape shown in Fig.~2.  In fact, 
for single-crystal ferromagnets the shape of the hysteresis loop is usually 
dependent on the orientation of the applied field with respect to the 
crystalline axes.  For materials composed of many different 
magnetic particles or grains, the nonlinear effects of each particle 
must be added to obtain a composite hysteresis loop.  
For such composite materials, or for bulk materials with impurities, 
the hysteresis loop is not smooth but contains small jumps, 
called Barkhausen jumps, that correspond to successive switching of 
small regions of the material (\BARKH).  
These loops can sometimes be parameterized 
using a model called the Preisach model (\PRIES).  

\centerline{Magnetic Energies}

The energies that are relevant to the 
properties of magnetic materials arise from a variety of physical effects.  
Which ones are most important for a particular 
engineering application depend on 
the composition of the particular piece of material, 
its mesoscopic structure (grain size, local stress, etc.), 
its surface properties, 
and its size and shape. 

\noindent\hfill{Exchange Energy}

The exchange energy comes from the quantum-mechanical 
overlap and hybridization of 
the exchange integrals between atoms and molecules.  
Since the interaction constant $J$ 
is due to the overlap of orbitals, it is a short-range 
interaction which often does not extend beyond nearest-neighbor pairs of 
lattice sites.  
For most materials $J$ is somewhat temperature 
and stress dependent, due to changes in interatomic distances.  
It is also dependent on the local environment of a particular magnetic 
atom.  Consequently exchange interactions for atoms at a surface, 
near a grain boundary, or near an impurity atom may be different 
from the exchange interaction of the same kind of atom in the bulk of 
a crystal.  
Often the Hamiltonian for a magnetic material is written as a 
sum of Heisenberg terms, such as ${\cal H}=J{\vec S}_i\cdot{\vec S}_j$,
where the three-dimensional spins ${\vec S}_i$ and ${\vec S}_j$ are 
located at nearest-neighbor sites $i$ and $j$ of the crystalline lattice.  

\noindent\hfill{Crystalline Anisotropy Energy}

When magnetic atoms are arranged on a crystal lattice, 
their spins have a lower energy if they are aligned along 
certain directions.  These are called the easy axes.  
The directions which require the highest energy 
for the orientation of the magnetic spins are the hard 
directions or hard axes.  Consequently, certain 
magnetization directions are preferred.  
One consequence of crystalline anisotropy is 
that hysteresis loops, such as the one shown in Fig.~2, 
depend on the angles between the applied magnetic field and the crystal axes.  
In magnetic materials which are 
not perfect crystals the orientation of the easy axes in 
different grains will most likely be different.  

For example, iron is a cubic crystal, and the three cube edges are the easy 
axes.  The anisotropy energy of Fe is expressed as 
$$
U_K = 
K_1\left(\alpha_1^2\alpha_2^2+\alpha_2^2\alpha_3^2+\alpha_3^2\alpha_1^2\right) 
+ K_2 \alpha_1^2\alpha_2^2\alpha_3^2
\eqno(4)
$$
where the $\alpha$'s are the direction cosines for the angles 
between the cube edges 
and the direction of the magnetization.  Here $K_1$ and $K_2$ 
depend on temperature and are zero above $T_{\rm c}$.  

\noindent\hfill{Dipole-Dipole Energy}

The classical interaction between two magnetic dipoles 
${\vec m}_1$ and ${\vec m}_2$ 
(which are vector quantities) is given by 
$
U({\vec m}_1,{\vec m}_2) = 
{{\mu_0}\over{4\pi r^3}} 
\left[{\vec m}_1\cdot{\vec m}_2 - 3\left(
{\hat r}\cdot{\vec m}_1\right)\left({\hat r}\cdot{\vec m}_2\right)\right] ,
$
where ${\vec r}=r{\hat r}$ is the 
vector from ${\vec m}_1$ to ${\vec m}_2$.  
This gives a long-range interaction between the dipoles.  
Since each spin in Fig.~1 represents a magnetic dipole, 
this dipole-dipole
interaction must be considered in dealing with magnetic particles.  
In many engineering applications 
where there are no time varying external fields, the microscopic motion 
of each individual spin 
(due to precession and random thermal fluctuations) need not be taken
into account, 
and the familiar equations of magnetostatics are recovered.
Consequently, the dipole-dipole interaction is sometimes called
the magnetostatic energy.  In particular, from magnetostatics 
the energy stored in the fields is given by 
$$
E_M=-{1\over 2}\int {\vec M}\cdot{\vec H} d\tau
\eqno(5)
$$ 
where the integral is over all ferromagnetic bodies.  
This may be understood 
as the integral of the interaction of each dipole with 
the field ${\vec H}$ created by all the other magnetic dipoles.  

In general the field inside a uniformly magnetized ferromagnetic 
body is not uniform, except in very special cases, such as 
general ellipsoids.  
For a general ellipsoid which is uniformly 
magnetized, the demagnetizing field inside the ellipsoid can be written as 
${\vec H}_{\rm d} = -{\bf N}{\vec M}$ where 
${\bf N}$ is a tensor.  
The demagnetizing field is due to the magnetic field caused by all 
the other dipoles in the particle.  
When ${\vec M}$ is parallel to one of 
the principal axes of the ellipsoid, the tensor is diagonal, 
and its three diagonal elements are called the demagnetizing factors.  
Then the magnetostatic self-energy of a uniformly 
magnetized ellipsoid of volume $V$ is given by 
$$
E_M = {V\over 2} \left(N_xM_x^2+N_yM_y^2+N_zM_z^2\right)
\eqno(6)
$$
since the tensor is diagonal.  
The demagnetizing factors may not be equal, so some directions 
will be preferred.  
This gives a shape anisotropy term in the energy of a ferromagnet.  
Note that the preferred directions from shape anisotropy can be 
different from the preferred directions for crystalline anisotropy.  
For more complicated geometries, a numerical method of 
solving for ${\vec H}_{\rm d}$ and evaluating the integral in Eq.~(5) 
is recommended.  

The dipole-dipole interaction is responsible for the formation of 
magnetic domains.  These domains are equilibrium regions 
where the magnetization is predominantly oriented in a single direction.  
They are separated by domain walls.  
Domain walls cost some energy due to the exchange and crystalline 
energy terms.  
However, this energy cost is balanced by the 
reduction in the magnetostatic energy resulting from the reduction in the 
total magnetization.  
Hence large magnetic particles break up 
into domains magnetized in different directions and separated by domain walls.  
Domain walls are classified 
according to the orientation of the spins in the domain-wall region.  

For small 
enough particles the energy cost of a domain wall is always higher than 
the gain resulting from the reduction in the 
magnetostatic energy. As a result, sufficiently small particles 
consist of a single magnetic domain. 
Nanometer-sized Co particles 
created by electron beam lithography may have only one or a few domains.  
Figure~3 shows pictures of an array of these particles.  Figure~3(a) 
shows the particles themselves, while Figure~3(b) shows the magnetic 
structure of the same particles.  
Even though all of the particles 
have similar sizes and histories, their domain structures are
different.  This illustrates both the history dependence of 
magnetic domains and the sensitivity of the magnetization to 
small changes in the geometry, environment, and composition 
of a magnetic particle.  

\noindent\hfill{Substrate and Surface Energies}

The surface of a magnetic particle can 
influence its physical behavior as much or even more 
than the bulk.  
This is because the energies associated with 
the exchange interactions and the crystalline anisotropy are 
extremely sensitive to the local environment.  
The environment of a magnetic atom is different at the surface 
and in the bulk.  
Surface effects become increasingly important as the particle size is 
decreased, due to the increased surface-to-volume ratio.  

Surface and interface effects can have important engineering applications.  
For example, giant magnetoresistive (GMR) materials 
can be 
grown by coupling a ferromagnetic film or particle to an antiferromagnetic 
bulk, leading to 
an exchange interaction between the spins of the two types of 
materials.  This gives an exchange bias which leads to an asymmetric 
hysteresis loop.  This is highly desirable 
because it enhances the signal-to-noise ratio of the read heads 
used in magnetic recording applications.  

Surfaces can also affect the response of a magnetic particle to changes in 
its environment.  For example, if the direction of the external field is 
reversed, reversal of the magnetization of the particle may be initiated 
at the surface.  This can make the coercive field of small particles 
significantly smaller than it would be in the absence of surfaces.  

\noindent\hfill{Strain}

The application of stress to a magnetic material results in 
a strain response.  The sensitivity of 
the local magnetization to the local environment means that strain 
in the lattice will change the local energies (the exchange energy, 
the crystalline anisotropy energy, and the dipole-dipole energy), 
which can alter the behavior of the magnetization locally.  

\noindent\hfill{Temperature}

The energy scales associated with the effects discussed above
must be compared with 
the thermal energy $k_{\rm B}T$, which causes the randomness of the spins 
seen in Fig.~1(d).  
Here $k_{\rm B}$ is Boltzmann's constant, 
and the temperature $T$ is given in Kelvin.  If the thermal energy 
is comparable to the other energies, it can affect engineering 
applications.  For instance, at nonzero temperature 
there is some probability that 
the orientation of the magnetization of a particle will 
change spontaneously.  This can lead to loss of data integrity 
in magnetic recording media ({\IEEEREV}). 

\centerline{Dynamics of Magnetic Particles}

The magnetization dynamics of magnetic particles 
are important for engineering applications. For example, consider a 
bit of information to be stored on one of the single-domain 
particles shown in Fig.~3.  The coding may be that 
a North-South orientation corresponds to Boolean 0 and a South-North 
orientation to Boolean 1.  Engineering questions 
involving the magnetization dynamics include: 
1) How strong a field must be applied to change a 0 to a 1 or vice versa?  
2) How long can a particle encoded with a bit be exposed to a stray field 
without losing data integrity, and how strong a stray field can be 
tolerated?  
3) If a bit is written today, how long can it be trusted to remain 
as it was written?  

\noindent\hfill{Metastability}

Arguably the most important concept needed to understand the dynamics of 
magnetic particles is metastability.  
One familiar example of this ubiquitous natural phenomenon is the 
supercooling of water.  
That metastability should be relevant to magnetism is 
suggested by the fact that the magnetization 
${\vec M}$ of a magnetic particle is history dependent.  

Figure~4 (a) shows the standard picture of metastability in a 
bistable system, such as a uniaxial single-domain magnetic particle.  
There are two free-energy minima, one of which is 
only a local minimum (the metastable state), and one of which is 
the global minimum (the equilibrium state).  
If the system is in the metastable state, 
a free-energy barrier, $\Delta {\cal F}$, must be overcome 
before the system can relax to the equilibrium state.  
The average lifetime of the metastable state, $\tau$, is given by
$$
\tau=\tau_0\exp\left({{\Delta{\cal F}}\over{k_{\rm B} T}}\right)
\eqno(7)
$$
where $\tau_0$ is an inverse attempt frequency.  In magnetic particles 
$\tau_0$ is typically taken to be on the order of  
an inverse phonon frequency, usually $\tau_0 \approx 10^{-10}$~sec.   
More detailed analysis shows that $\tau_0$ depends on the 
curvature of the free energy at both the metastable 
minimum and the saddle point (the maximum 
that separates the metastable and equilibrium states).  

The system can get into the metastable state due to its history.  
How can it get out?  
There are two possible answers to this question.  
Think in analogy to a 
pitcher half-filled with water, which is held above a sink.  The 
water is the system, the pitcher is the metastable state, and the sink is the 
equilibrium state.  One way to get the water into the sink 
is to shake the pitcher to make the water 
splash out.  The more vigorously it is shaken, the 
faster the water will splash out.  This method of escape from a 
metastable state is analogous to a magnetic material 
at a nonzero temperature.  As seen from Eq.~(7), the lifetime 
increases as the height of the barrier increases, whereas it decreases as 
the temperature increases. 
Another way of getting the water into the sink is to 
tip the pitcher.  As the pitcher continues to tip, at some point the 
water starts to flow out.  This corresponds to 
an escape from the metastable state of a magnetic particle, which 
is valid even at zero temperature.  In particular, the 
barrier $\Delta {\cal F}$ depends on the applied field 
${\vec H}_{\rm appl}$, and changing ${\vec H}_{\rm appl}$ corresponds 
to tipping the pitcher.  
At a particular value of the field $\Delta 
{\cal F}$ equals zero, and the metastable state disappears.  
This situation is illustrated in Fig.~4(b).  The field that must be applied 
for $\Delta {\cal F}$ to vanish is often called the nucleation field, 
$H_{\rm nucl}$.  

This pitcher analogy also illustrates an important consideration concerning 
the escape from a multidimensional metastable well at zero temperature.  
In particular, how far the pitcher must be tipped before the water 
starts to spill out depends on the direction in which it 
is tipped.  The smallest angle is needed 
if the pitcher is tipped to make the water spill from the spout --- the 
lowest part of the rim. 
Similarly, the nucleation field, $H_{\rm nucl}$, 
often depends on the direction of the applied field.  
However, for thermally driven escape the decay always proceeds across 
the saddle point (analogous to the spout) 
as the temperature (analogous to the amplitude of the shaking) is increased, 
or as the waiting time is increased at fixed temperature. 

The waiting time is related to 
another physical quantity of interest, 
the probability that a magnetic particle starting in the metastable state 
at $t$$=$$0$ has never left it at time $t$. 
This probability is often called $P_{\rm not}(t)$. 
It is only in the last decade that it has become possible 
to measure $P_{\rm not}(t)$ for individual single-domain magnetic 
particles ({\EXPRA},{\EXPRB}). For thermal 
escape over a single barrier one has 
$$
P_{\rm not}(t) = \exp\left( - {{t} / {\tau}}\right)
,
\eqno(8)
$$
where $\tau$ is given by Eq.~(7).  

Equation~(7) is the fundamental equation needed to understand the dynamics 
of a magnetic particle.  
All one requires is a knowledge of $\Delta{\cal F}$.  
This, however, requires a knowledge of the energy of the spin 
configuration at the saddle point.  Since $\Delta{\cal F}$ 
enters in an exponential, the value of $\tau$ is extremely sensitive 
to small changes in $\Delta{\cal F}$.  

\noindent\hfill{Coherent Rotation}

Consider a spherical single-domain uniaxial magnetic particle 
with crystalline anisotropy constant $K$ 
and a uniform magnetization ${\vec M}$ 
in an applied magnetic field along the easy axis.  If all the spins 
in the particle point in the same direction, the total energy 
is $E=KV\sin^2(\theta)-M_{\rm s} V H\cos(\theta)$.   Here 
the saturation magnetization $M_{\rm s}$ makes an angle $\theta$ with the 
easy axis, and the volume of the particle is $V$.  
In this case it is possible to obtain the zero-temperature free-energy 
barrier exactly, namely 
$\Delta{\cal F}=K V\left[1+\left({{H M_{\rm s}}\over{2K}}\right)^2\right]$.
In this model all the spins rotate coherently and act like one large 
spin.  
Consequently, particles that behave this way are called superparamagnetic 
particles.  
Note that the zero-temperature energy barrier has been 
assumed to be valid at finite temperatures in this model 
of a superparamagnet. 
Since superparamagnetism implies escape over a single barrier,
$P_{\rm not}(t)$ is given by Eq.~(8).  

The particle volume $V$ enters the metastable lifetime in an 
exponential, so a small change in particle size can lead to extremely 
large changes in $\tau$.  For example, for an iron sphere 
with a radius of $115$~{\AA} the lifetime $\tau\approx 0.07$~s.  
If the radius is $140$~{\AA}, then 
$\tau\approx1.5\times10^5$~s ({\AHARO}).  

The coherent rotation mode described above is often called the 
N{\'e}el-Brown reversal mode ({\AHARO}) 
since N{\'e}el derived Eq.~(7) with 
$\Delta{\cal F}$$=$$KV/k_{\rm B}T$ and Brown wrote a differential 
equation for a random walk in the metastable well to obtain a 
non-constant prefactor for Eq.~(7).  
It is also possible to obtain a 
zero-temperature hysteresis curve using the same 
assumptions as above, namely a uniform magnetization 
and that only the external field and a uniaxial anisotropy are 
important.  This is called the Stoner-Wohlfarth model ({\AHARO}).  
This model gives an upper bound for the coercive field 
as $H_{\rm c}\le 2K/M_{\rm s}$.  

An equivalent analysis for the case 
in which shape anisotropy is important can also be performed.  
Again, with the assumption that all the 
spins always point in the same direction, the analysis is the same 
except that $K$ now arises from shape anisotropy.  
The assumption that the spin configuration at the saddle point 
has all spins pointing in the same direction  
is only sometimes valid.  There are other zero-temperature reversal 
modes where the zero-temperature 
saddle point has spins in other configurations.  Examples 
include modes descriptively named buckling, curling, and fanning.  
The dominant mode depends on the geometry 
and size of the magnetic particle. 

\noindent\hfill{Nucleation and Growth}

As the volume $V$ increases, 
the average rate for magnetization reversal via coherent rotation 
quickly becomes too small to be practically important.  
This was illustrated by the example with iron particles discussed above. 
Other reversal modes with lower free-energy barriers can then come into 
play, especially for highly anisotropic materials, in which domain 
walls are relatively thin. 

At nonzero temperatures, thermal fluctuations continually create 
and destroy 
small ``droplets'' of spins aligned with the applied field. The free energy of 
such a droplet 
consists of two competing parts: a positive part due to the interface 
between the droplet and the metastable background, 
and a negative part due to the alignment of the spins in the 
droplet with the field. For a droplet of radius $R$, these terms are 
proportional to $R^{d-1}$ and $- |H| R^d$, respectively. 
Here $d$ is the spatial dimension of the particle, and these relations hold 
both for three-dimensional particles ($d$$=$$3$) and 
for particles made of a thin film ($d$$=$$2$).  
Small droplets most likely shrink.  
However, if a droplet manages to become larger 
than a critical radius, $R_{\rm c} \propto 1/|H|$, it will 
most likely continue to grow and eventually bring the whole particle into the 
stable magnetization state. The free-energy barrier associated with such 
a critical droplet is proportional to $1/|H|^{d-1}$, 
which is {\it independent\/} of the particle size! 
Since the droplets can only grow at a finite speed, in sufficiently 
large particles or for sufficiently strong fields,
new critical droplets may nucleate at different positions while the 
first droplets are still growing. 

The droplet switching mechanism sketched above gives rise to three regimes of 
field strengths and particle sizes: \hfil\break
1: For sufficiently weak fields and/or small particles, a critical droplet 
would be larger than the entire particle. As a result, the saddle point 
configuration consists of an interface 
which cuts across the particle, so that  
$\Delta {\cal F} \propto V^{(d-1)/d}$, independent of $|H|$ to lowest order. 
The behavior in this regime is essentially superparamagnetic, even though 
the dependence of $\Delta {\cal F}$ on the particle size 
is somewhat weaker than predicted for 
uniform rotation.  
In this regime both Eq.~(7) and Eq.~(8) are valid.  \hfil\break
2: For stronger fields and/or larger particles, the magnetization 
reverses by the action of a single droplet of the stable phase. 
The free-energy barrier is independent of the particle volume, but 
because the droplet can nucleate anywhere in the particle, the average 
lifetime is inversely proportional to $V$. 
We call this decay regime the Single-Droplet regime. 
A series of snapshots of a computer simulation of Single-Droplet 
decay is shown in Fig.~5. 
In this regime Eq.~(8) is still valid.  
\hfil\break
3: For yet stronger fields and/or larger particles, the decay occurs via a 
large number of nucleating and growing droplets. In this regime, 
which we call the Multi-Droplet regime, the average lifetime is 
independent of $V$. 
A series of snapshots of a computer simulation of Multi-Droplet 
decay is shown in Fig.~6. 
In this regime Eq.~(8) is no longer valid, and $P_{\rm not}(t)$ 
takes the form of an error function (11,12).  \hfil\break

\noindent\hfill{Switching Fields}

Two rather similar quantities that are 
often measured for magnetic particles are the switching field, 
$H_{\rm sw}$, and the coercive field, $H_{\rm c}$. The former is 
defined as the magnitude of the 
field for which a particle switches with probability 1/2 
within a given waiting time after field reversal. 
The latter is the value of the sinusoidal field at which the 
hysteresis loop crosses the field axis, as shown in Fig.~2. Both 
depend 
weakly 
on the time scale of the experiment (waiting time or inverse field 
frequency), but they are qualitatively similar over a wide range 
of time scales. A collection of experimentally measured coercive fields for 
various materials are shown as functions of the particle size in Fig.~7(a). 
The increase in $H_{\rm c}$ for small sizes is due to the 
superparamagnetic behavior of the particles, while the 
decrease for larger sizes is due to the dipole-dipole 
interaction, which causes large particles to 
break up into multiple domains.  
For particles in the nanometer range, which are single-domain in 
equilibrium, the crossovers between the 
three nucleation driven magnetization reversal 
mechanisms described above give rise to very similar 
size dependences ({\HOWAR},{\GROUP}), as seen in Fig.~7(b).

\noindent\hfill{Domain Boundary Movement}

If the magnetic particle is multidomain in zero field, the application of 
a magnetic field will cause the domain wall(s) to move.  In this case 
the dynamics of the magnetization is dominated by the domain-wall movement.  
Typically there will be pinning sites due to impurities, grains, and 
surfaces, that the domain wall must overcome before it can move.  These 
obstacles can either be overcome by applying a sufficiently large field, 
or by waiting for the random thermal fluctuations to move the domain 
wall past the pinning centers.  This is analogous to the 
zero-temperature and finite-temperature reversal mechanisms in 
single-domain particles.  
If the domain wall moves due to random thermal fluctuations the 
magnetization of the particle will change slowly with time, 
a phenomenon called magnetization creep (12).  

\noindent\hfill{Magnetic Viscosity}

Consider a large number of identical non-interacting particles.  
If a strong field is applied and then removed, the remanent 
magnetization will decay with time as 
$
M_{\rm r}(t)=M_{\rm r}(0) \exp\left(-t/\tau\right)
$
as particles cross the barrier separating the two equilibrium states.  
However, in the typical case the particles are not identical, and 
there is a distribution of lifetimes, ${\cal P}(\tau)$.  
In this case the time decay is
$$
M_{\rm r}(t)=M_{\rm r}(0) \int_0^\infty {\cal P}(\tau)\exp(-t/\tau) d\tau
\eqno(9)
$$
Under some circumstances and for certain specific distributions 
for ${\cal P}(\tau)$ Eq.~(9) can be approximated by 
$
M_{\rm r}(t)\approx C-S\ln(t/{\bar t}_0)
$
where $C$, $S$, and ${\bar t}_0$ are constants.  This 
logarithmic decay of the magnetization is called magnetic viscosity.  
It must be emphasized that this logarithmic equation 
is a valid approximation only under certain specific circumstances 
and even then only for a limited range of $t$ ({\AHARO}).  

\centerline{Magnetocaloric Effects}

Magnetocaloric effects of magnetic particles have engineering 
applications principally in refrigeration.  
Current applications are mostly to ultra-low temperature 
refrigeration, but near-term applications to higher temperatures 
seem promising.  
Given that at constant volume the $PdV$ work term is zero, 
the first law of thermodynamics becomes
$$
{\dbar}Q = dU - {\dbar}w=dU - \mu_0{\vec H}\cdot {\dbar}{\vec M}
\eqno(10)
$$
where $U$ is the internal energy of the magnetic particle 
and the symbol ${\dbar}$ denotes an infinitesimal change rather than 
a true differential.  This must be considered because of the 
history dependence of ${\vec M}$.  
For an adiabatic change, ${\dbar}Q=TdS=0$, where $S$ is the entropy.  
Then the relation
$$
\Delta T = -{T\over C_H} 
\left({{\partial M}\over{\partial T}}\right)_H \Delta H
\eqno(11)
$$
can be obtained ({\CRAIK}), 
where $C_H$ is the heat capacity at constant field.  
Thus an increase in $H$ produces a rise in temperature, and 
vice versa, which is the magnetocaloric effect.  
This can be utilized as a cooling mechanism by placing the 
magnetic material in a strong applied field, and then turning off 
the applied field.  This will lead to a decrease in temperature 
due to adiabatic demagnetization according to Eq.~(11).  

\centerline{Acknowledgments}

Work supported by
the U.S.\ National Science Foundation Grant No.\ DMR-9520325, 
and by Florida State University 
through the Center for Materials Research and Technology and 
the Supercomputer Computations Research Institute 
(U.S.\ Department of Energy Contract No.\ DE-FC05-85ER25000).

%
%

\vfill
\eject

\centerline{Bibliography}

{\EXPRA}.~~M.\ Lederman, S.\ Schultz, and M.\ Ozaki, 
Phys.\ Rev.\ Lett.\ {\bf 73},  1986 (1994).  

{\EXPRB}.~~W.\ Wernsdorfer, E.\ Bonet Orozco, K.\ Hasselbach, A.\ Benoit, 
B.\ Barbara, N.\ Demoncy, A.\ Loiseau, H.\ Pascard, and D.\ Mailly, 
Phys.\ Rev.\ Lett.\ {\bf 78}, 1791 (1997).   

{\EXPRC}.~~C.\ Salling, S.\ Schultz, I.\ McFadyen, and M.\ Ozaki, 
IEEE Trans.\ Magn.\ {\bf 27},  5184  (1991); 
C.\ Salling, R.\ O'Barr, S.\ Schultz, I.\ McFadyen, and M.\ Ozaki, 
J.\ Appl.\ Phys.\ {\bf 75}, 7989 (1994).

{\EXPRD}.~~V.\ Cros, S.~F.\ Lee, G.\ Faini, A.\ Cornette, A.\ Hamzic, 
and A.\ Fert, J.\ Magn.\ Magn.\ Mater.\ {\bf 165},  512  (1997). 

{\SCOTT}.~~S.~W.\ Sides, P.~A.\ Rikvold, and M.~A.\ Novotny, 
J.\ Appl.\ Phys.\ {\bf 83}, June (1998).  

{\STEIN}.~~C.~P.\ Steinmetz, 
Trans.\ Am.\ Inst.\ Electr.\ Eng.\ {\bf 9}, 3 (1892).

{\BARKH}.~~O.\ Perkovic, K.\ Dahmen, J.~P.\ Sethna, 
Phys.\ Rev.\ Lett.\ {\bf 75}, 4528 (1995).  

{\PRIES}.~~I.~D.\ Mayergoyz, {\it Mathematical Models of Hysteresis\/}, 
New York, Springer, 1991; 
C.~E.\ Korman and P.\ Rugkwamsook, IEEE Trans.\ Magn.\ {\bf 33}, 
4176 (1997).  

{\IEEEREV}.~~T.\ Yogi and T.~A.\ Nguyen, 
IEEE Trans.\ Magn.\ {\bf 29}, 307 (1993); 
P.~L.\ Lu and S.\ Charap, IEEE Trans.\ Magn.\ {\bf 30}, 4230 (1994); 
I.\ Klik, Y.~D.\ Yao, and C.~R.\ Chang, 
IEEE Trans.\ Magn.\ {\bf 34}, 358 (1998);  
J.~J.\ Lu and H.~L.\ Huang, 
IEEE Trans.\ Magn.\ {\bf 34}, 384 (1998);  
E.~N.\ Abarra and T.\ Suzuki, 
IEEE Trans.\ Magn.\ {\bf 33}, 2995 (1997);  
Y.\ Hosoe, I.\ Tamai, K.\ Tanahashi, Y.\ Takahashi, T.\ Yamamoto,
T.\ Kanbe, and Y.\ Yajima, 
IEEE Trans.\ Magn.\ {\bf 33}, 3028 (1997).  

{\AHARO}.~~A.\ Aharoni, {\it Introduction to the Theory of Ferromagnetism}, 
Oxford, Oxford University Press, 1996. Ch.~5. 

{\HOWAR}.~~H.~L.\ Richards, S.~W.\ Sides, M.~A.\ Novotny, and P.~A.\ Rikvold, 
J.\ Magn.\ Magn.\ Mater.\ {\bf 150}, 37 (1995). 

{\GROUP}.~~P.~A.\ Rikvold, M.~A.\ Novotny, M.\ Kolesik, and H.~L.\ Richards in 
{\it Dynamical Properties of Unconventional Magnetic Systems\/}, 
NATO Science Series E: Applied Sciences, Vol.~349, 
edited by A.~T.\ Skjeltorp and D.~Sherrington, 
Kluwer, Dordrecht, 1998. 

{\TEBBL}.~~R.~S.\ Tebble and D.~J.\ Craik, {\it Magnetic Materials\/}, 
New York, John Wiley \& Sons, 1969.  

{\CRAIK}.~~D.\ Craik, 
{\it Magnetism, Principles and Applications}, 
New York, John Wiley \& Sons, 1995; Chapter~1.   

%
%

\vfill
\eject

\centerline{Reading List}

\noindent 1.~~A.\ Aharoni, {\it Introduction to the Theory of Ferromagnetism}, 
Oxford, Oxford University Press, 1996.  

\noindent 2.~~H.~N.\ Bertram, {\it Theory of Magnetic Recording}, 
Cambridge University Press, Cambridge, 1994.  

\noindent 3.~~S.\ Chikazumi, {\it Physics of Magnetism}, 
Malabar, Florida, Krieger Publishing Company, 1978.  

\noindent 4.~~D.\ Craik, 
{\it Magnetism, Principles and Applications}, 
New York, John Wiley \& Sons, 1995.  

\noindent 5.~~J.~C.\ Mallinson, {\it The Foundations of Magnetic Recording}, 
Second Edition, 
Boston, Academic Press, 1993.  

\noindent 6.~~For a review of recent experimental studies of magnetization 
switching in magnetic nanoparticles, see the introduction of: 
H.~L.\ Richards, M.\ Kolesik, P.-A.\ Lindg{\aa}rd, P.~A.\ Rikvold, and 
M.~A.\ Novotny, Phys.\ Rev.\ B {\bf 55}, 11521 (1997).

\vfill
\eject

\centerline{Figure Captions}

\noindent 
Fig.~1.  Schematic representation of the arrangement of 
local magnetic moments or spins in 
different types of magnetic materials.  
In (a) all the spins are 
aligned, and the material is a ferromagnet.  
The total magnetization, given by the sum of all the spin vectors, 
is large.  
In (b) the nearest-neighbor spins are all anti-aligned, so the 
total magnetization is zero.  This arrangement is that of 
an antiferromagnetic material.  
In (c) the nearest-neighbor spins are also anti-aligned,
but here the spins on each sublattice have different lengths.  
Hence the total magnetic moment is nonzero in this ferrimagnetic material.  
In (d) the directions of the 
spins are randomly distributed due to thermal fluctuations, 
so that the 
total magnetic moment is zero.  This represents a paramagnet.  
Above a critical temperature, $T_{\rm c}$ 
(which is different for different materials), all magnetic materials 
become paramagnetic.  
 
\noindent 
Fig.~2.  
A hysteresis loop for a model of a single-domain 
uniaxial magnetic particle, the Ising model.  
An external magnetic field $H_0\sin(\omega t)$ is 
applied, and the magnetization is recorded as a function of time.  
Shown on the loop is the location of the average coercive field, $H_{\rm c}$.  
This is the field which must be applied to make the 
magnetization equal to zero.  
Also shown is the remanent, or spontaneous, 
magnetization, $M_{\rm r}$.  This is the magnetization when the applied 
field is equal to zero.
In this particular model the value of $M_{\rm r}$ is known exactly.  
The saturation magnetization, $M_{\rm s}$, 
corresponds to all the spins 
aligned, as in Fig.~1(a).  
The fluctuations on the hysteresis loop are due to random thermal noise, 
which is typically seen in hysteresis loops for small particles 
at nonzero temperatures.  
This loop is for a temperature of $0.97T_{\rm c}$.  
The loop would become smoother at lower temperatures and for larger 
particles than the 64$\times$64 lattice used here.  
The loop area can be a complicated function of $H_0$, $\omega$, 
and temperature (\SCOTT).  
The data used to generate this figure are from a Monte Carlo 
simulation, courtesy of Dr.\ Scott W.\ Sides.  

\noindent 
Fig.~3.  
Nanometer-sized magnetic particles of cobalt (Co).  These were grown 
as a polycrystalline film on 
a GaAs substrate, and electron beam (e-beam) lithography was used to 
make isolated particles. 
The substrate in this case 
should contribute minimally to the magnetization 
of the particles.  
The size of these particles are 
roughly 170~nm, which is about as 
small as can currently be manufactured using e-beam lithography.  
Although the particles look uniform, as seen in the 
Atomic Force Microscope (AFM) image (a), due to the history dependence 
of magnetic materials different magnetizations are seen in 
the Magnetic Force Microscope (MFM) image (b).  This image measures 
the magnetic field outside the magnetic particles caused by 
their individual magnetizations.  
The dark region corresponds to the North pole 
and the light region corresponds to the South pole.  
Some of the particles are seen 
to be single domain, while others have a more complicated 
magnetic arrangement.  
These are the domain patterns for the islands after manufacture, before 
any magnetic field was applied.  
They are schematically illustrated in 
the sketches (c) and (d).  
Thanks to Prof.\ Andrew Kent of the 
New York University physics department for the AFM and MFM images.  

\noindent 
Fig.~4.  
A metastable well is illustrated in (a).  Even though the 
lowest-energy state is on the right, an energy barrier 
$\Delta{\cal F}$ must be overcome before 
the metastable state on the left can decay to the equilibrium state.  
As the external magnetic field is increased, the 
barrier decreases, and it becomes zero at a particular field 
called the nucleation field, $H_{\rm nucl}$, as illustrated by 
the heavy curve in (b).  There is no metastable state for higher 
fields, as illustrated by the light curve in (b).  

\noindent 
Fig.~5.
Three snapshots from a computer simulation of the single-droplet 
switching mechanism for a model of a particle made from 
a highly anisotropic, uniaxial ultrathin magnetic film. 
Time increases from top to bottom in the figure. 
Data courtesy of Dr.\ S.~W.\ Sides. 

\noindent 
Fig.~6.
Three snapshots from a computer simulation of the multi-droplet 
switching mechanism for a model of a particle made from 
a highly anisotropic, uniaxial ultrathin magnetic film. 
Time increases from top to bottom in the 
figure. 
Reproduced from ({\HOWAR}) with permission from Elsevier Science B.V. 

\noindent 
Fig.~7.  
Plots of the coercive and switching fields are shown versus particle size.  
As explained in the text, these fields can be 
considered to be roughly the same.  
(a) The coercive field, $H_{\rm c}$,  
for magnetic particles of various materials, shown versus the particle 
diameter.  
Reproduced from ({\TEBBL}) with permission from John Wiley \& Sons, Ltd.  
(b) For a simple model for a uniaxial single-domain particle, 
the $L$$\times$$L$ two-dimensional Ising model, 
simulations show a maximum in $H_{\rm sw}$ (in units of $J$) 
versus $L$, even when 
there are no dipole-dipole interactions so that particles 
of all sizes remain single domain.  
This effect is due to different nucleation decay mechanisms for 
particles of different size ({\HOWAR},{\GROUP}), as described in the text.  

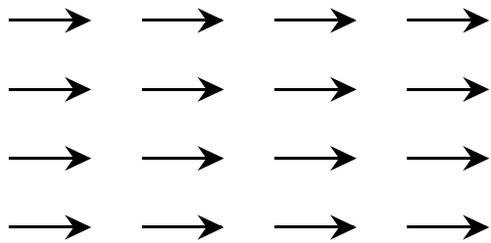
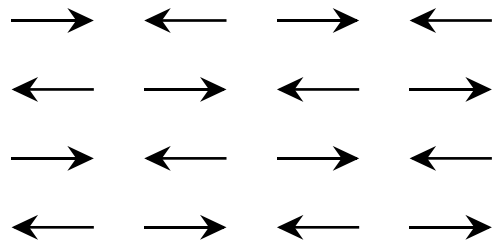
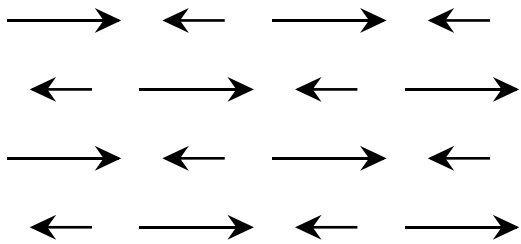
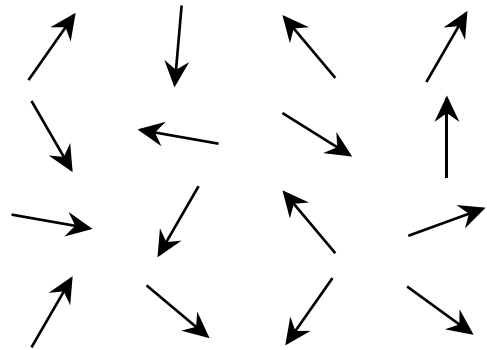
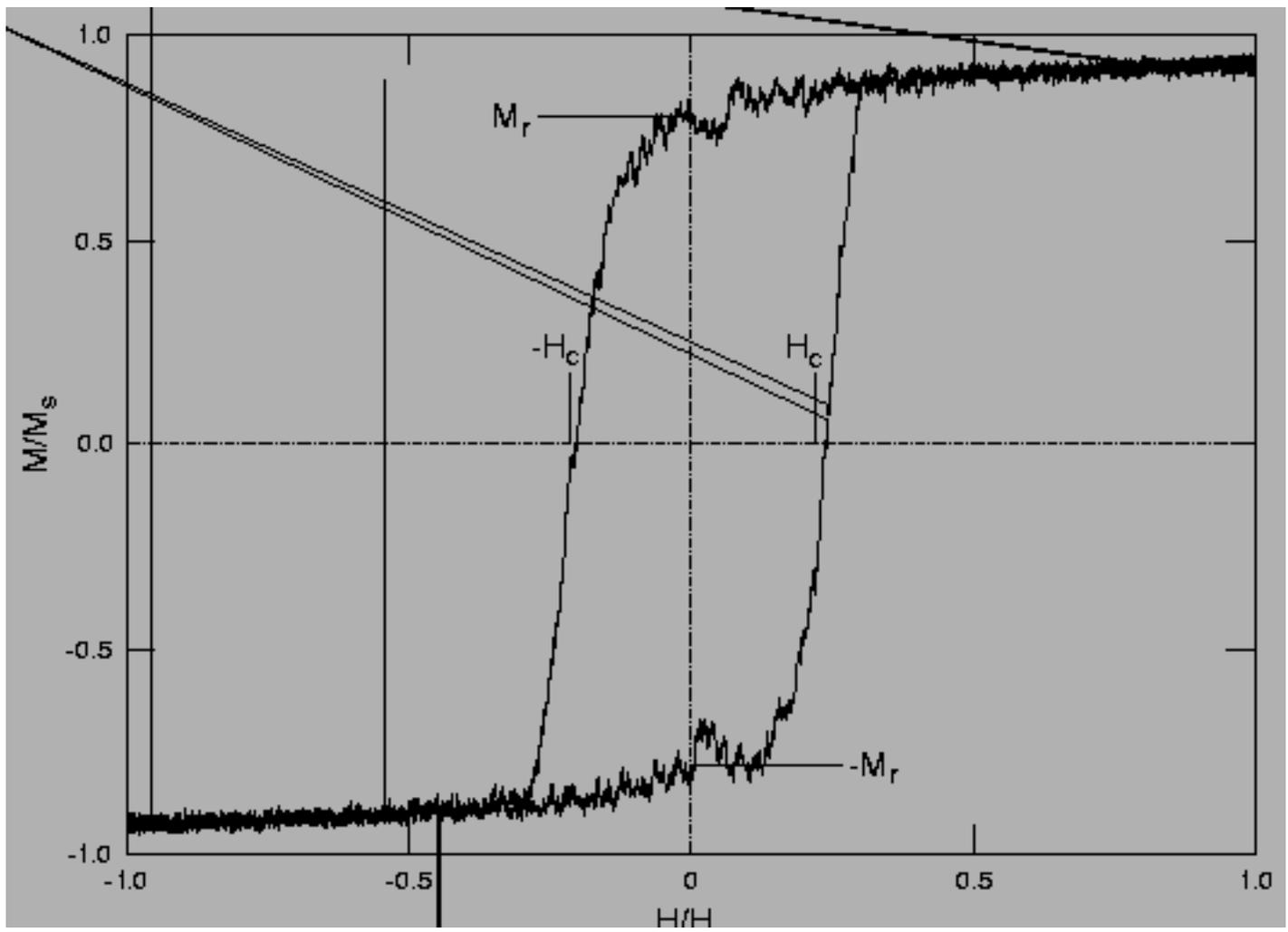
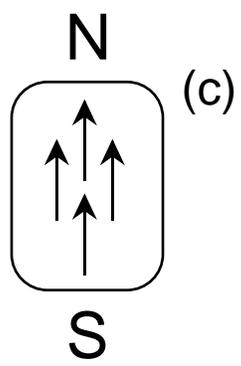
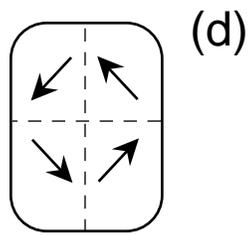
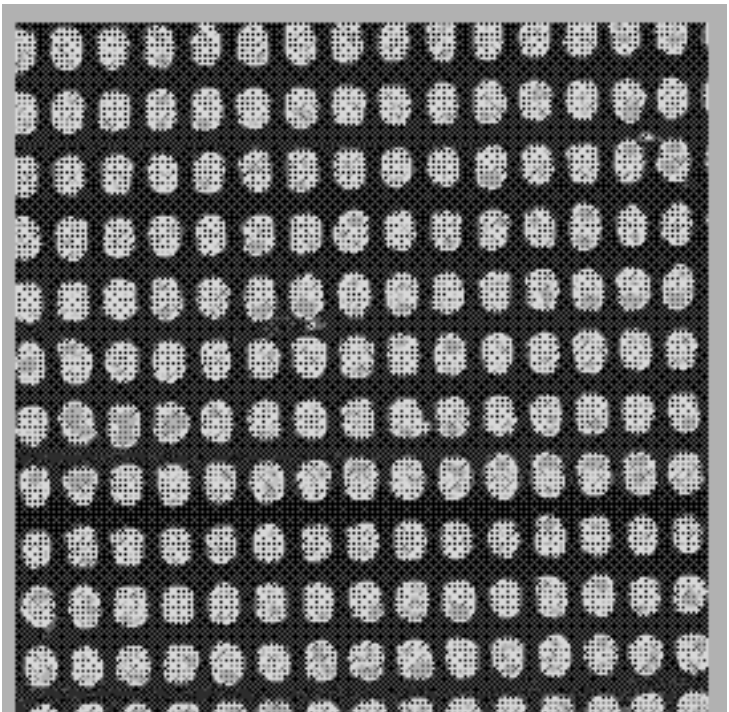
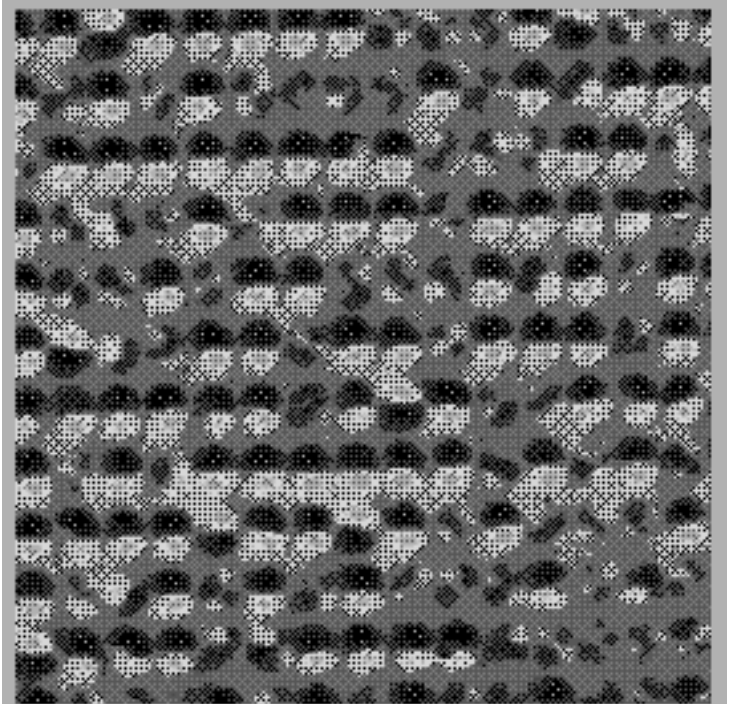
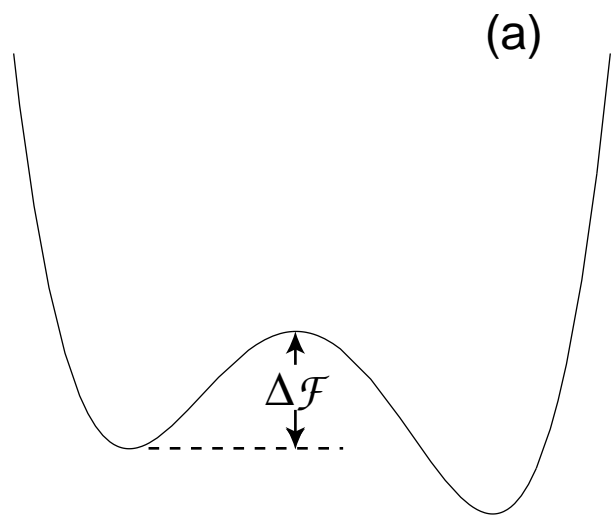
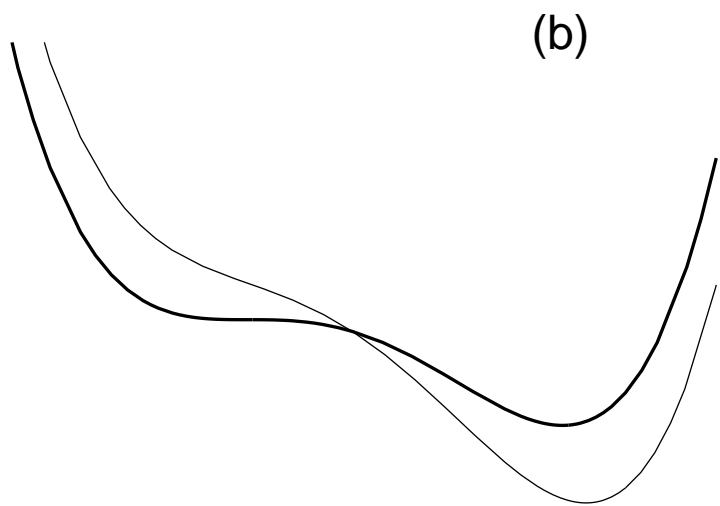
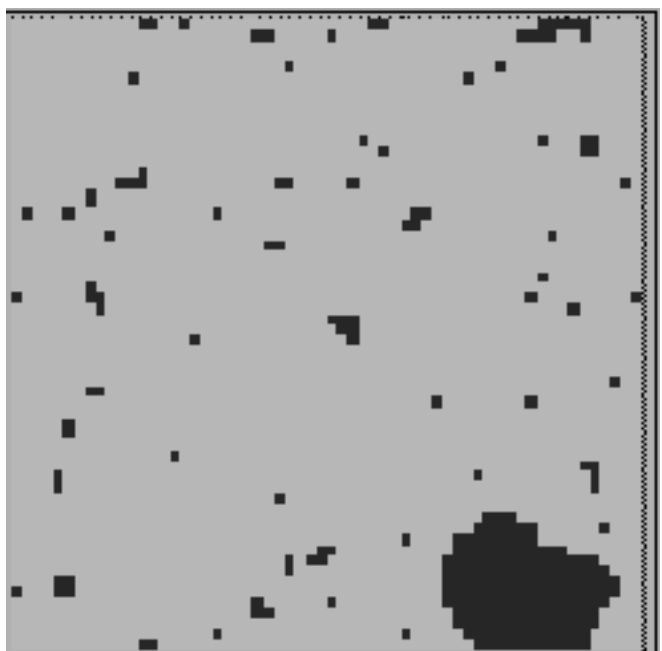
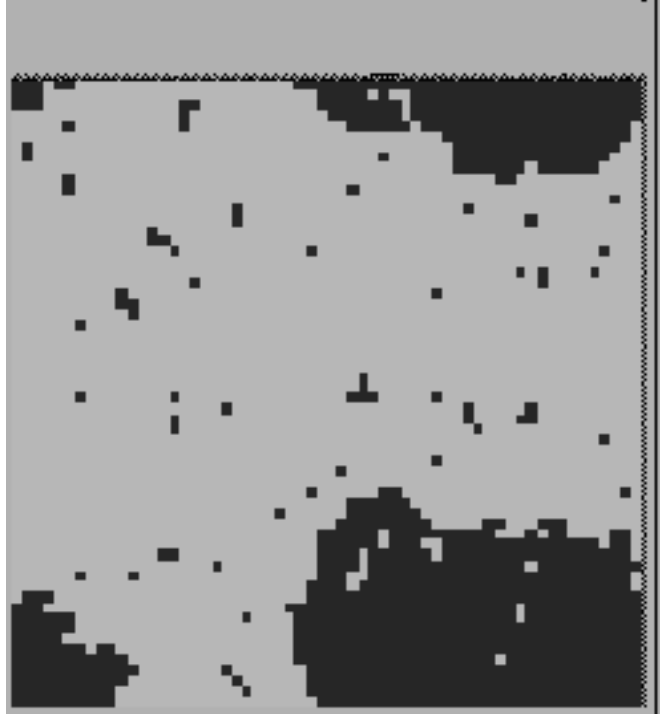
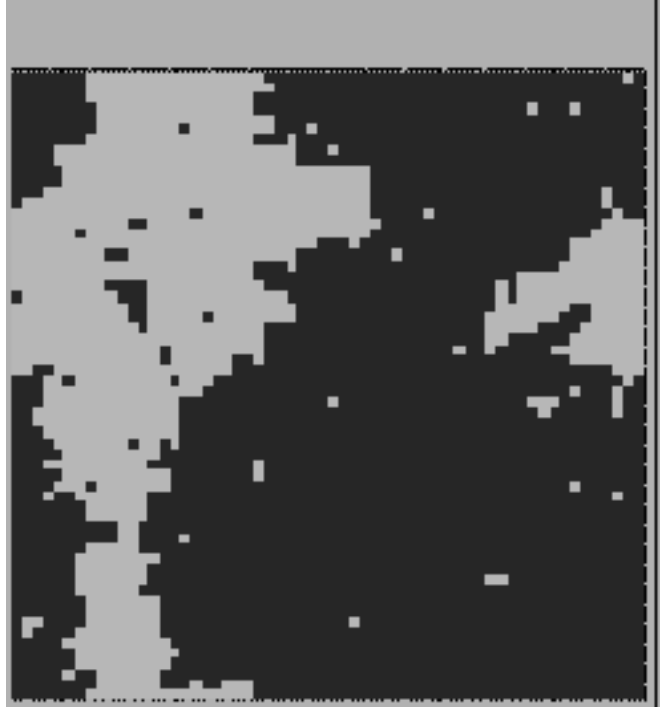
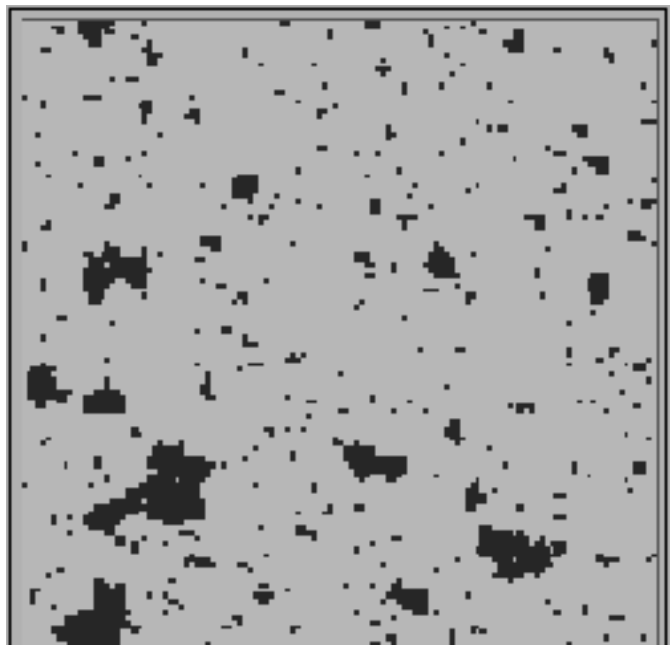
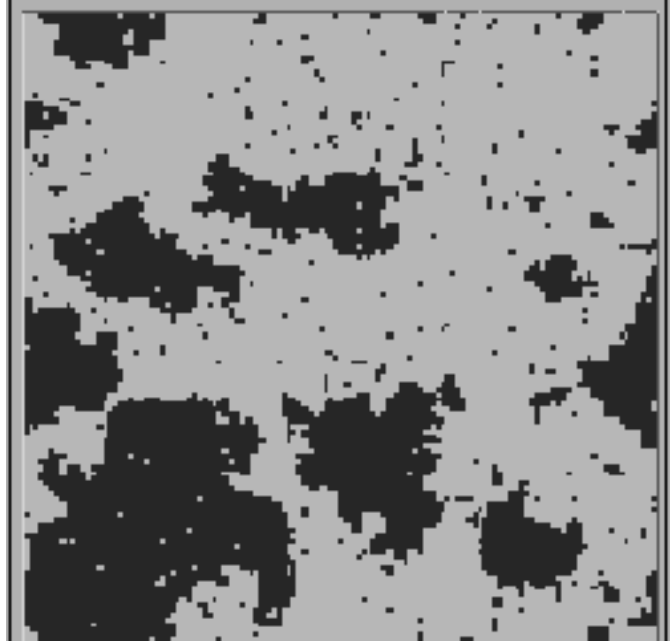
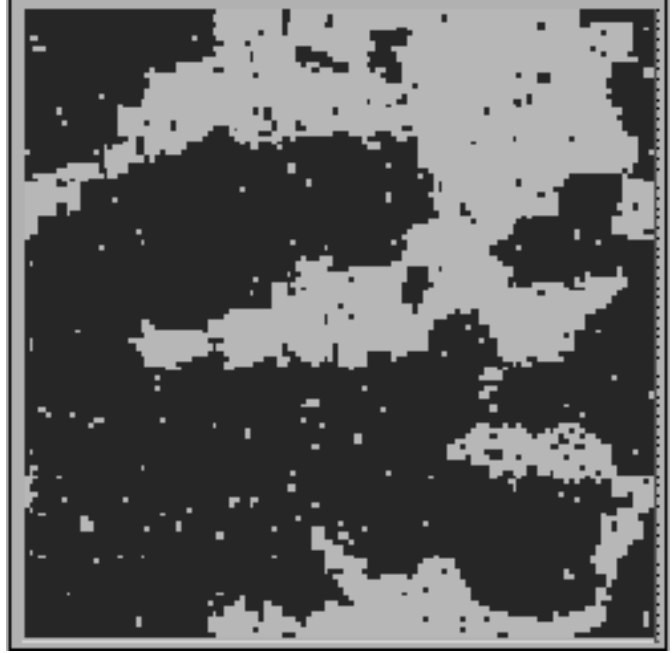
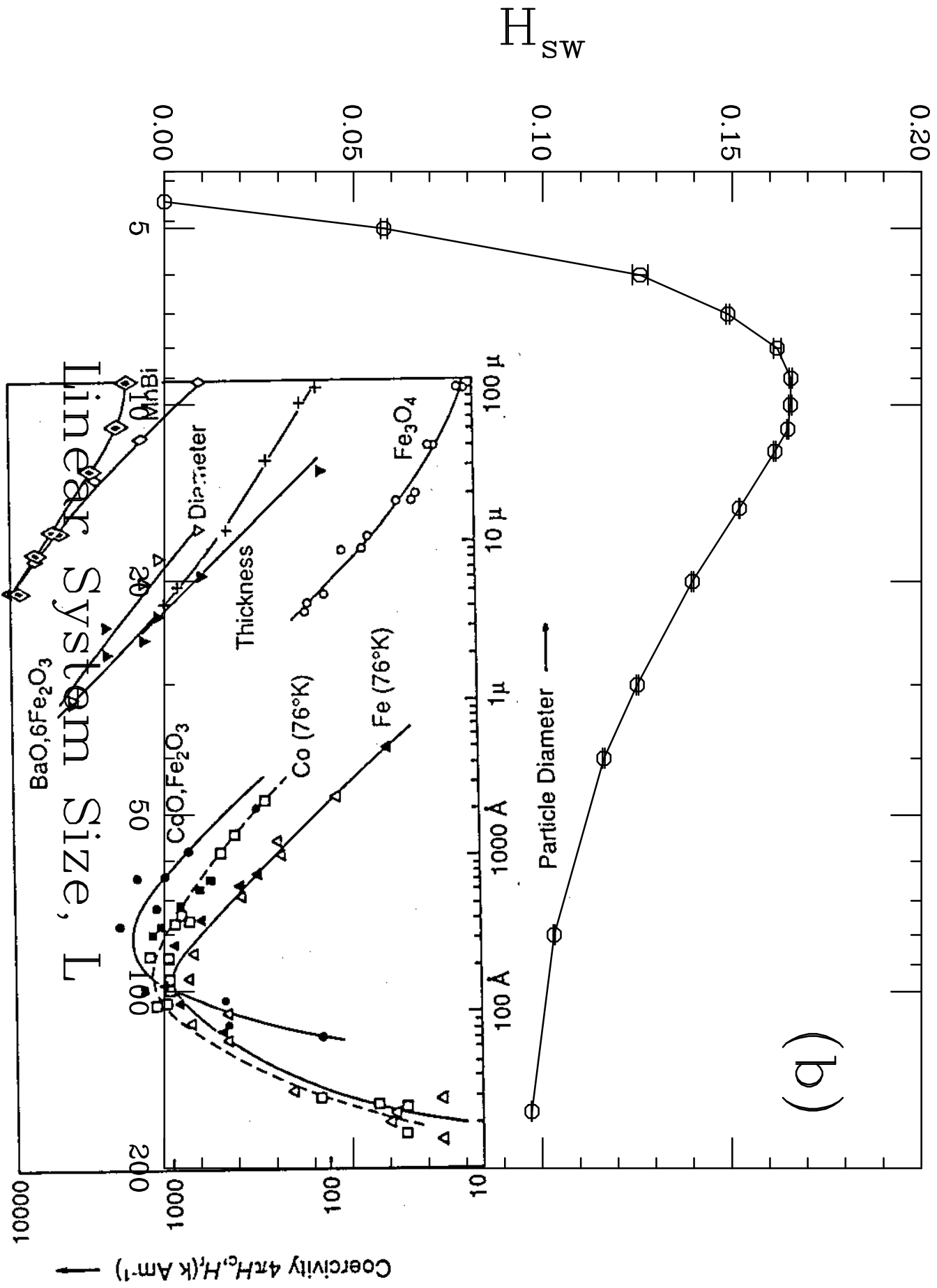
\bye